# Recent Advances in Energy Efficient Resource Management Techniques in Cloud Computing Environments

July 13, 2021

**Niloofar Gholipour**[1] ,**Ehsan Arianyan**[2]  , **Rajkumar  Buyya**[3]

1  ICT Research Institute, Tehran,Iran; n.gholipour@itrc.ac.ir
2  ICT Research Institute, Tehran,Iran;  ehsan_arianyan@itrc.ac.ir
3  University of Melbourne, Australia; rbuyya@unimelb.edu.au

**Abstract**

Nowadays cloud computing adoption as a form of hosted application and services is widespread due to decreasing costs of hardware, software, and maintenance. Cloud enables access to a shared pool of virtual resources hosted in large energy-hungry data centers for diverse information and communication services with dynamic workloads. The huge energy consumption of cloud data centers results in high electricity bills as well as emission of a large amount of carbon dioxide gas. Needless to say, efficient resource management in cloud environments has become one of the most important priorities of cloud providers and consequently has increased the interest of researchers to propose novel energy saving solutions. This chapter presents a scientific and taxonomic survey of recent energy efficient cloud resource management' solutions in cloud environments. The main objective of this study is to propose a novel complete taxonomy for energy-efficient cloud resource management solutions, review recent research advancements in this area, classify the existing techniques based on our proposed taxonomy, and open up new research directions. Besides, it reviews and surveys the literature in the range of 2015 through 2021 in the subject of energy- efficient cloud resource management techniques and maps them to its proposed taxonomy, which unveils novel research directions and facilitates the conduction of future researches.

## 1  Introduction

Cloud computing is an internet-based paradigm with the capability of delivering service model including infrastructure, platform, and software as services, which are made available as subscription-based services under the circumstance of a pay-as-you-go model to consumers.  In other words, there is a pool of computing resources that consumers can access them on-demand[1]. There are varieties of definitions of cloud computing but the U.S National Institute of Standards and Technology  (NIST) in its definition of cloud computing [2] describes "a "Measured Service" as being one of the five essential characteristics of  the  cloud  computing  model. Providing data on measurable capabilities (such as; quality of service, security features, availability and reliability) gives the cloud service customer the opportunity to make informed choices and to gain understanding of the state of the service being delivered.  It also gives the cloud service provider the opportunity to present the properties of their cloud services to the cloud service customer".

Some of the important features of cloud computing are on-demand self-service provisioning, broad network access necessity, availability of resource pools, multi-tenancy, and elasticity.

As depicted in figure 1, there are three main cloud service models: Infrastructure as a service (IaaS), Platform as a service (PaaS), and Software as a service (SaaS) [3].

In addition to traditional cloud services, there is a new type of cloud service called CaaS that has been introduced recently [4]. A particular example of container management system can be mentioned as Docker that permits developers to define containers for their application and develop, run and manage their applications without any concern about the underlying infrastructure and required software [5]. Containers share the same host operating system kernel, and they have communication with physical hardware via system call that is much faster than the traditional communication-based hypervisor. Accordingly, they are defined as lightweight virtual environments compared to virtual machines that provide an isolation layer between workloads without the overhead of hypervisor-based virtualization. A Container as a building block of CaaS cloud model suggests an isolated virtual environment without any monitoring media such as hypervisor that is existed in traditional systems [4]. Containerization increases the efficiency of resource utilization compared to virtual machines. Google container engine and Amazon Elastic Container Service (Amazon ECS) currently supply this new method.

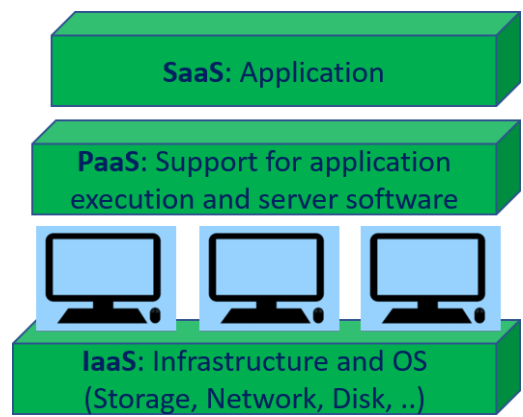

Figure 1: Cloud service model

## 1.1 Motivation and contribution

With the rapid ever-increasing demand to access diverse information and communication technology services based on the cloud service delivery model, the number of huge energy-hungry cloud data centers is increasing rapidly. Thus, these days energy consumption in cloud environments is a Crucial issue. A data center will consume the energy of about 1000TWH in next ten years (2013-2025) [6]. The percentage of energy consumption by the data centers and cooling systems will reach 5% of the total energy consumption in the world. Energy consumption leads to operational cost and environmental implications such as global warming [7].
This key challenge leads to a rethink about the techniques and research strategies to lessen energy consumption as a crucial matter in the cloud environment. To overwhelm this challenge, there are various solutions that researchers have introduced; among them, resource management techniques play a significant role. Nonetheless, energy efficiency is still a challenge for future researchers [8]. Virtualization technique enable cloud providers to create multiple Virtual Machine(VM) instances on a single physical server, or multiple containers on a VM or a physical server which makes it possible to have servers with higher utilization. The dynamic consolidation of both VMs and containers through live migration is an efficient approaches for saving energy in cloud data centers [9]. Although the recent technological developments and paradigms including High Performance Computing (HPC), containerization, exascale computing, and processing at network edge appear to yield new opportunities for cloud computing, they are also creating new challenges and demands for new approaches and research strategies.
Container technology has emerged thanks to Docker [10] which has boosted in both academia and industry. It provides a way to package an application that can be run with its dependencies and libraries isolated from other applications. Containers arose as a lightweight alternative of VMs that present better microservice architecture supports. The technology of container is strongly supported by PaaS, IaaS, and Internet Service Providers. Traditional hypervisor-based solutions are virtualized at the hardware level, while containerization provides virtualization at the operating system level. The containers interact with each other via system standard calls and they do not have any information about themselves [11]. Although VM technology needs to have an individual operating system for each VM, only one operating system can serve all containers in container technology. So, container technology provides more lightweight virtual systems which makes it possible to utilize system resources such as CPU, RAM, and network bandwidth more efficiently [10]. This happens owing to Linux kernel's cgroups and namespaces which are used by docker. Besides, utilizing container technology considerably decreases startup time and the expected resources for each image in comparison with VM technology. To exemplify, a container requires 50 milliseconds to start, while a VM is activated in 30-40 seconds [12].

Many Internet companies have embraced this technology and containers have become the de-facto standard for creating, publishing, and running applications. On the other side, there are still impediments and challenges in container-based virtualization demanding to be addressed, including security issues, in particular during migration, dynamic resource allocation, and energy consumption [13].

Container orchestration appears to yield several possibilities for cloud and application providers to determine how to select, deploy, monitor, and dynamically manage the configuration of containers in the cloud. It is also involved with the management at runtime to support the deploy, run, and maintain states [14]. As a consequence, an especially serious problem to address in this context is the scheduling or placement of containerized applications on the available hosts along with VM consolidation. Our main contributions are as follows:

- Addressing an Energy-aware cloud resource management problem.

- Proposing taxonomies for energy aware resource management techniques in the cloud environment.

- Studying and reviewing the important issues and parameters in the literature for energy aware cloud resource management approaches.

- Proposing open research directions related to energy-aware resource management problem in cloud environment.

The aim of this chapter is to express the importance of energy-efficient resource management in cloud environments and present a scientific and taxonomic survey of the relevant recent literature in the period of 2015 through 2021. We propose a novel taxonomy for the subject of energy-efficient resource management in cloud environments and categorize the most recent papers utilizing our proposed taxonomy. In this chapter, we address nine other survey papers in the literature as our related work which are about energy-efficient resource management. We also evaluate the cons and pros of these papers and compare them with our survey based on our proposed taxonomy. In addition, we study and compare 32 articles based on our proposed taxonomy.

Our proposed taxonomy appears to yield a number of distinctions in comparison with other related surveys. One of the most significant superiority is considering the containerization technology and container migration topic in addition to virtualization and VM migration techniques that previous works neglected. Unlike other similar taxonomies that consider the dynamic voltage and frequency scaling (DVFS) technique applied only on Central processing unit(CPU), our taxonomy addresses the DVFS approach applied at two levels: memory and CPU. Other relevant papers consider only arbitrary workloads in their work, while our taxonomy considers different types of workloads, including high-performance computing (HPC), real-time, as well as batch applications. Another differences between this chapter and previous studies can be mentioned as reflecting both passive and active resource types. Finally, our taxonomy considers rack and geographically level consolidations as well as server-level consolidation. The last important superiority of our chapter is that it surveys the latest state-of-the-art papers in the literature.

The rest of the chapter is organized as follows: Section 2 reviews the related works, followed by our proposed taxonomy for energy-aware resource management in cloud environments and virtualization in Section 3 and 4, respectively. Comparing and mapping the recent relevant papers based on our proposed taxonomy is fulfilled in section 5. Ultimately, future research directions and conclusions, are presented in section 6 and 7.

## 2 Related Work

Energy-efficient resource management approaches in cloud environments is a hot topic which vastly addressed by researchers. Since cloud computing's research has advanced continuously, there is a need for a systematic review to evaluate, update and join the existing literature. This section summarizes some of the previous works in the literature similar to our work.

The authors in [15] have conducted a survey on energy-aware resource allocation in cloud data centers. They have reviewed some keywords such as virtualization, allocation of VM, energy efficiency, power consumption, as well as cloud computing. They have discussed various kinds of energy-aware system architectures for the cloud, and compared energy efficiency in both traditional and virtual data centers. This chater has further proposed a taxonomy for energy-saving methods in cloud data centers, which were studied in three levels, such as power management, resource management, and thermal management. The researchers have reviewed previous works based on VM allocation algorithms, VM selection algorithms, and Dynamic Voltage Frequency Scaling(DVFS), which conduced to energy-saving. Plus, they have shown that the energy-saving approach became possible using renewable energy that plenty of recent research introduced this strategy.

In [16] authors have presented a brief survey describing primary energy-conserving techniques in the cloud environment. To add, they have classified energy consumption approaches into five categories, including energy-efficient hardware, energy-aware scheduling, consolidation, energy conservation in a cluster of servers, as well as power-efficient networks. Finally, they have evaluated a few papers based on this classification. The researchers further have focused on consolidation techniques in three levels, containing task consolidation, server consolidation, and energy-aware task consolidation.

Researchers in [17] have performed a comprehensive survey on energy-efficient computing, clusters, grids, and clouds. They have reported a number of approaches in the literature which contributed to improve energy efficiency. This chapter has proposed three taxonomies, covering such levels as scheduling, energy efficient computing, as well as energy-efficient technique at different levels to make data center greener. Plus, [17] studied the energy efficiency of a single system and large-scale cloud data centers, storage systems, and networking.

Scientists in [18] have reviewed energy-saving strategies in computational clouds. They explored state-of-the-art related to energy efficiency as well as performance administration, vitality for effective data centers, and resource distributions. What is more, they have studied the existing techniques in four stages, including tools, OS, virtualization, and data center. Their proposed taxonomy was divided into static and dynamic power management at the highest level. On the one hand, They have considered DVFS, resource throttling, Dynamic Component Deactivation (DCD), and workload consolidation at the software level in their taxonomy. On the other hand, they have considered Dynamic Performance Scaling (DPS), resource throttling, DVFS, and DCD at the hardware level.

The authors in [19] have carried out a survey on several papers over the last decade respecting energy efficiency techniques for cloud computing applications. They have mentioned some approaches consisting of heuristic algorithms for live VM migration, task scheduling, load balancing. This paper has further discussed various kinds of tools that are applied in order to design energy efficient schemes. They have analyzed three simulators in this scope, including CloudSim, a popular tool that supports the IaaS layer, GreenCloud, and icanCloud. In [23] scientists have conducted a survey on energy-aware VM consolidation strategies, highlighting the limitations and profits. They have analyzed VM consolidation features such as power consumption, VM consolidation types, and so on. Besides, they have compared state-of-the-art VM consolidation techniques in terms of objectives, Service Level Agreement (SLA) violation, energy cost, number of VM migrations, etc. To add, they have categorized these proposed techniques in a table; however, they have not presented any taxonomy.

Table 1: Comparison table

| Author | Year | Providing Taxonomy | Considering Passive Resources | Hybrid Approach | Geographical level in consolidation | Workload type | | | DVFS | | Migration type | |
|---|---|---|---|---|---|---|---|---|---|---|---|---|
| | | | | | | Batch | Real-time | HPC | CPU | Memory | Hybrid | CR/TR |
| N.Akhter et al. [15] | 2016 | ✓ | ✓ | | ✓ | | | | ✓ | | | |
| S.Singh et al. [16] | 2016 | | ✓ | | | ✓ | ✓ | | | | | |
| S.Kaur et al. [20] | 2016 | ✓ | ✓ | | | ✓ | | | | | | |
| M.Zakarya. [21] | 2016 | ✓ | ✓ | | ✓ | | ✓ | ✓ | ✓ | | | |
| Q.Shaheen et al. [18] | 2018 | ✓ | | ✓ | | ✓ | ✓ | ✓ | | | | |
| D.Hazra et al. [22] | 2018 | | | ✓ | ✓ | | ✓ | ✓ | ✓ | | | |
| N.Hamdi et al. [23] | 2019 | | | | | | | | | | | |
| S.Puhan et al. [19] | 2020 | | ✓ | | | | | | NM | NM | | |
| Q.Zhou et al. [24] | 2020 | | ✓ | ✓ | | | | | NM | NM | | |
| Existing survey | 2021 | ✓ | ✓ | ✓ | ✓ | ✓ | ✓ | ✓ | ✓ | ✓ | ✓ | ✓ |

The researchers in [24] introduced a survey on energy efficient algorithms based on VM consolidation for Cloud Computing. In this paper, five state-of-the-art energy-efficient algorithms were compared from architecture, modeling, and metrics points of view; Modified Best Fit Decreasing (MBFD), EcoCloud, Greedy based scheduling Algorithm Minimizing Total Energy (GRANITE), Learning Automata Overload Detection (LOAD), as well as Ant Colony System (ACS). What is more, they have implemented and analyzed these algorithms according to the experimental settings in the CloudSim simulator using the PlanetLab workload. As their results have shown, ACS can achieve the best energy efficiency in most cases as it also has the least number of active hosts.

Another survey which was addressed by [20] has concentrated on energy-aware VM placement and consolidation techniques in cloud environments. Researchers in this paper, have investigated such approaches as heuristic, constraint, and bin packing problem, so forth. To exemplify, they have considered some procedures, containing first fit, single-dimensional best fit, volume-based best fit, plus dot product based fit. In the same way, they reviewed several papers in their proposed categories. According to them, some significant issues that should be considered in this area pertains to workload consolidation, Quality of Service (QoS) guarantee, minimizing the number of both migrations and active physical hosts.

Researchers in [22] have briefly reviewed energy-aware task scheduling algorithms in cloud environments.They also have analyzed the detriments and benefits of existing approaches. One of the algorithms that they considered in their paper can be mentioned as an energy-aware genetic algorithm. Another considered approach can be referred to as an energy-saving task scheduling depending on vacation queue. In addition to these methods, they have studied the DVFS technique. To sum up, they concluded that most state-of-the-art papers do not notice the deadline of a task and the cost of executing a task while making the scheduling decisions.

## 3 Proposed taxonomy for energy-aware resource management solutions in cloud environments

In this section, we present our proposed taxonomy for energy-aware resource management solutions in cloud environments, as shown in Fig2. In our proposed taxonomy, we consider four items at the highest level. The first level pertains to the goals of energy-efficient resource management in cloud environments. As can be seen, the second level goes back to the dynamism of resource management technique. The third level is the considered type of workload, including arbitrary, High Performance Computing (HPC), batch, and real-time applications. Finally, the fourth level is the type of resources that are classified into active and passive. The details of this taxonomy are described in the following subsection.

## 3.1 Goals

As energy-efficient resource management approaches play a crucial role in cloud data centers these days, there are various kinds of goals in this way. To clarify, fig 2 summarises the components of goals. Indeed, we have considered five targets for this component, such as minimizing power consumption, maximizing performance, load balancing, meeting power budget, plus maximizing business profit. We have also regarded four performance metrics, including response time, SLA violation, throughput, and delay. First and foremost, data centers have significant power cost; thereby, it is an essential requirement for data centers' operation to meet the power budget coming from the limitation for power usage and observing this limitation [25].

Due to load imbalance, some of the data center resources may become overloaded or underloaded, which leads to performance degradation and resource wastage. Load balancing conduces to maximize resource utilization and achieve the desired QoS in the cloud by employing optimal resource allocation and workload distribution approaches at both schedule and run time.

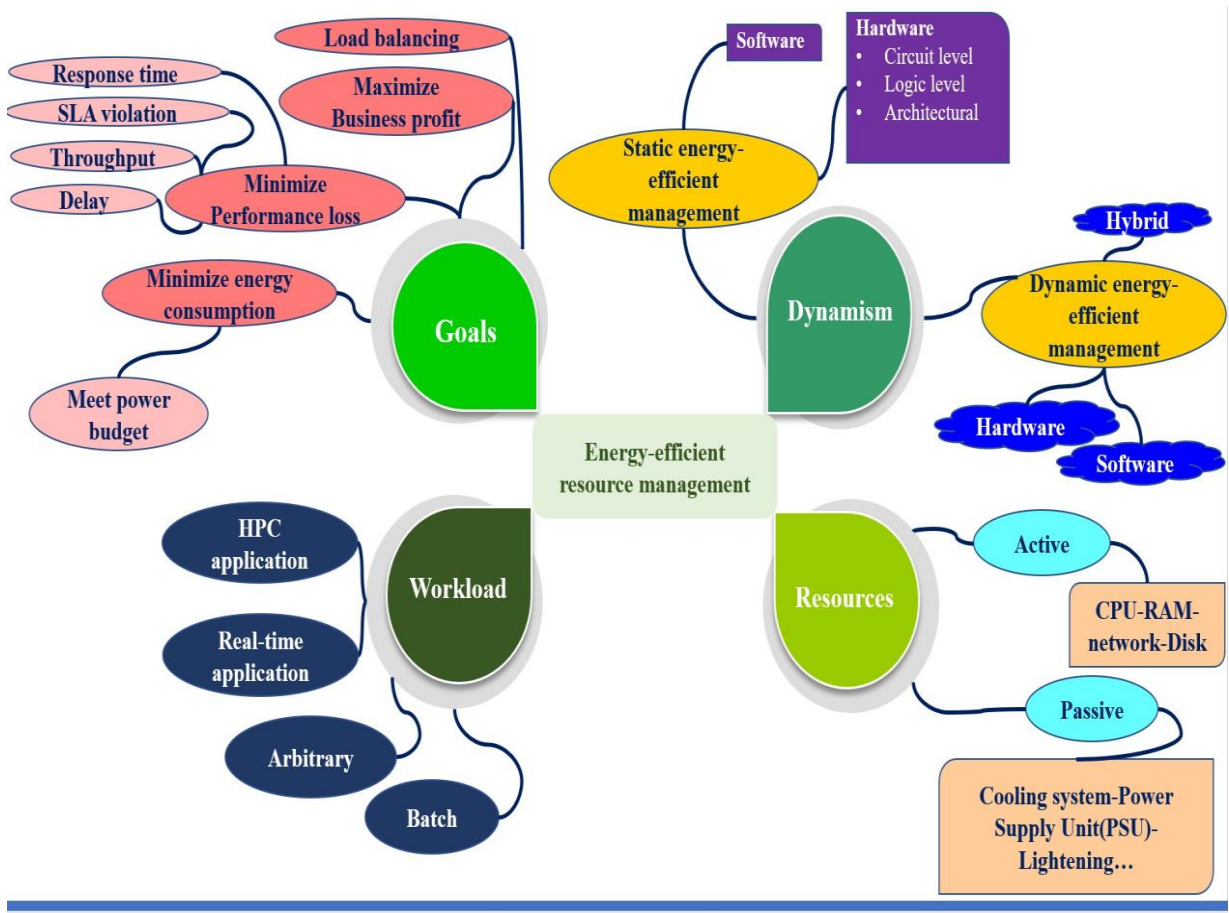

Figure 2: Taxonomy of energy efficient management solution in cloud environments

## 3.2 Workload

As a basic definition, the workload is a job that is characterized by release time, worst-case execution time, and deadline. At a high level, the workload is a sequence of jobs and tasks. The computer workload is defined as the amount of work allocated to the system that should be completed in a determined time [4]. A usual system workload comprises of tasks and user's requests submitted to the data center. As stated in [26], understanding the workload is far more important than designing new scheduling algorithms. If the systems are not tested using the correct input's workloads, the proposed policies or algorithms' results might not work as expected when applied to real-world scenarios [4]. We consider four workload types in our proposed taxonomy containing arbitrary, batch, HPC, and real-time workload, described as follows:

- **Batch processing**: Theoretically, batch processing is a processing mode when a sequence of jobs are executed on a batch of inputs [27]. Analyzing data on a large scale and batch processing occurs by utilizing data centers and some distributed and computing frameworks such as Map-Reduce and Hadoop [4]. Map-Reduce programming paradigm is the most practical and efficient solution for batch processing of big data [28].

- **HPC**: In the early 1990s, clusters of computers became famous in HPC environments owing to their low cost compared to custom supercomputers and mainframes [29]. Computers with high processing power and fast network connections. Also, HPC computers generally take advantage of open source operating systems such as Linux. In the early 2000s, grid computing was linked to the HPC community as a consequence of need to run parallel programs even larger than that was normal in grid environments. Grids provide powerful resources operated by independent administrative domains to users [30]. In the late 2000s, cloud computing was quickly growing its adolescent level and reputation, and studies started to appear on the viability of executing HPC applications on remote cloud resources [31], [32]. HPC applications are resource-intensive scientific workflows (in terms of data, computation, and communication) that have usually aimed at Grids and customary HPC platforms like super-computing clusters [33]. Both the size and number of HPC data centers have overgrown in recent years, which conduces to an exponential increase in power drastically [34].

- **Real-time application**: With improved cloud computing infrastructure, real-time computing can be accomplished on cloud infrastructure [35]. In most of the real-time cloud applications, the processing is executed on remote cloud computing nodes. As we meet many real-time systems around us, Cloud's support plays a crucial role in the real-time system [36]. Their application ranges from small mobile phones to huge industrial controls and from a mini pacemaker to larger nuclear plants. A usual real-time, such as financial analysis, distributed databases, or image processing, includes multiple real-time applications or subtasks service [37]. Real-time systems are implemented by several simultaneous tasks requesting to access hardware resources [17].

### 3.3 Resources

In cloud computing, resource management plays a key role in the entire system's performance [38]. Ordinarily, a data center comprises four main structural components including network switches, cooling systems, racks, also servers [39], [40]. Each rack consists of several servers, while each server has both a dedicated power unit and a cooling fan. In our proposed taxonomy, we categorize the resources into active and passive types. CPU, RAM, Disk, and network interface are considered as main active resources. Also, cooling system and power supply unit (PSU) are considered as main passive resource type.

### 3.4 Dynamism

Dynamism specifies the dynamicity of the power-aware resource management's techniques in our proposed taxonomy. From dynamism point of view, energy-aware resource management techniques are divided into static power management (SPM) and dynamic power management (DPM). The next subsections present these techniques in detail.

#### 3.4.1 Static power management (SPM)

The static power management can be executed on both hardware and software levels. Leakage currents in any active circuits cause static power consumption at the hardware level [7]. SPM uses hardware components such as CPU, memory, disk storage, network devices, and power supply unit efficiently.

It consists of all applied optimization methods during design time at logic, circuit, and architectural levels that will be explained in the following section.

- **Logic level optimization**: At this level, optimization methods attempt to optimize the power of switching activity in both sequential and combinational circuits. Minimizing the switching capacitance improves the dynamic power consumption straightly by reducing the energy per transition on each logic device [7], [41].

- **Circuit level optimization**: Significant challenges at this optimization level are based on efficient pipelining and interconnections between stages and components. Pipelining technique is regularly used to boost throughput in high-performance designs at the expense of reducing energy efficiency, contributing to increasing area and execution time [41].

- **Architectural level optimization**: Methods include the system's design considering power optimization technique at an architectural level [7]. Power savings are typically accomplished at the architectural level by optimizing the system components' balance to prevent wasting power.[41].

Besides the optimization at the hardware level, considering the SPM at the software level is also essential. Even with robust hardware design, it is crucial to be careful about software design inasmuch as weak design conduces to loss of power and performance, even with perfectly designed hardware. Thus, the code generation, the instructions used in the code, and the order of these instructions must be carefully selected, as they affect performance as well as power consumption.

### 3.4.2 Dynamic Power Management (DPM)

This section describes our taxonomy at the dynamic power management level, as shown in Fig. 3. DPM is categorized into three levels, including hardware, software, and hybrid. There are various kinds of optimization methods at both the hardware and software levels [7], [42]. At the hardware level, we can imply techniques such as DVFS, DCD, and sleep states. In addition, the techniques at the software level are classified into virtualization, migration, consolidation, plus containerization. The dynamic power consumption is induced by the high usage of hardware components (such as CPU, storage, and network devices) and the circuits' activity. The main reason enabling dynamic power consumption pertain to both system's components deactivation and tuning the circuit activity. Dynamic power consumption can be reached through different techniques including: 1) diminishing the switching activity, 2) decreasing the physical capacitance that relies on low-level design parameters such as transistors' sizes, 3) ebbing the supply voltage, and 4) lessening the clock frequency [7].
DPM improves energy consumption by using knowledge gathered from current resources in the system and the workload of applications running in the system [7], [43]. DPM techniques allow dynamic adjustment of power states to occur based on current system loads. It predicts the best action in the future using the data obtained from the system and according to the system's requirements. DPM techniques are categorized into hardware and software levels. There is another level in our taxonomy, namely hybrid, in which both hardware and software techniques are simultaneously utilized.

- **Hardware-level approaches**
  DPM techniques applied at the hardware level reconfigure the system dynamically by adopting methodologies to fulfill the requested services with the minimum number of active components or the minimum load on such components [43]. The DPM techniques at a hardware level can optionally turn off the idle system components or reduce the useless ones' performance. It is also possible to exchange some components, containing CPU, between either active or idle modes to save energy. The hardware DPM techniques vary for different hardware components, yet usually, they are splited into dynamic component deactivation (DCD) and dynamic performance scaling (DPS) [44].

  **Dynamic component deactivation (DCD)**
  The techniques in our proposed taxonomy at the DCD level are categorized into both partial dynamic system deactivation (PDSD) and complete dynamic system deactivation (CDSD) [43]. PDSD techniques are built upon the idea of the clock gating parts of an electronic component. Computer components, which do not support performance scaling and can be deactivated, need some techniques that leverage the unsustainable workload and disable the idle devices. Nonetheless, CDSD techniques are based on the idea that the components are entirely deactivated during some periods [43].

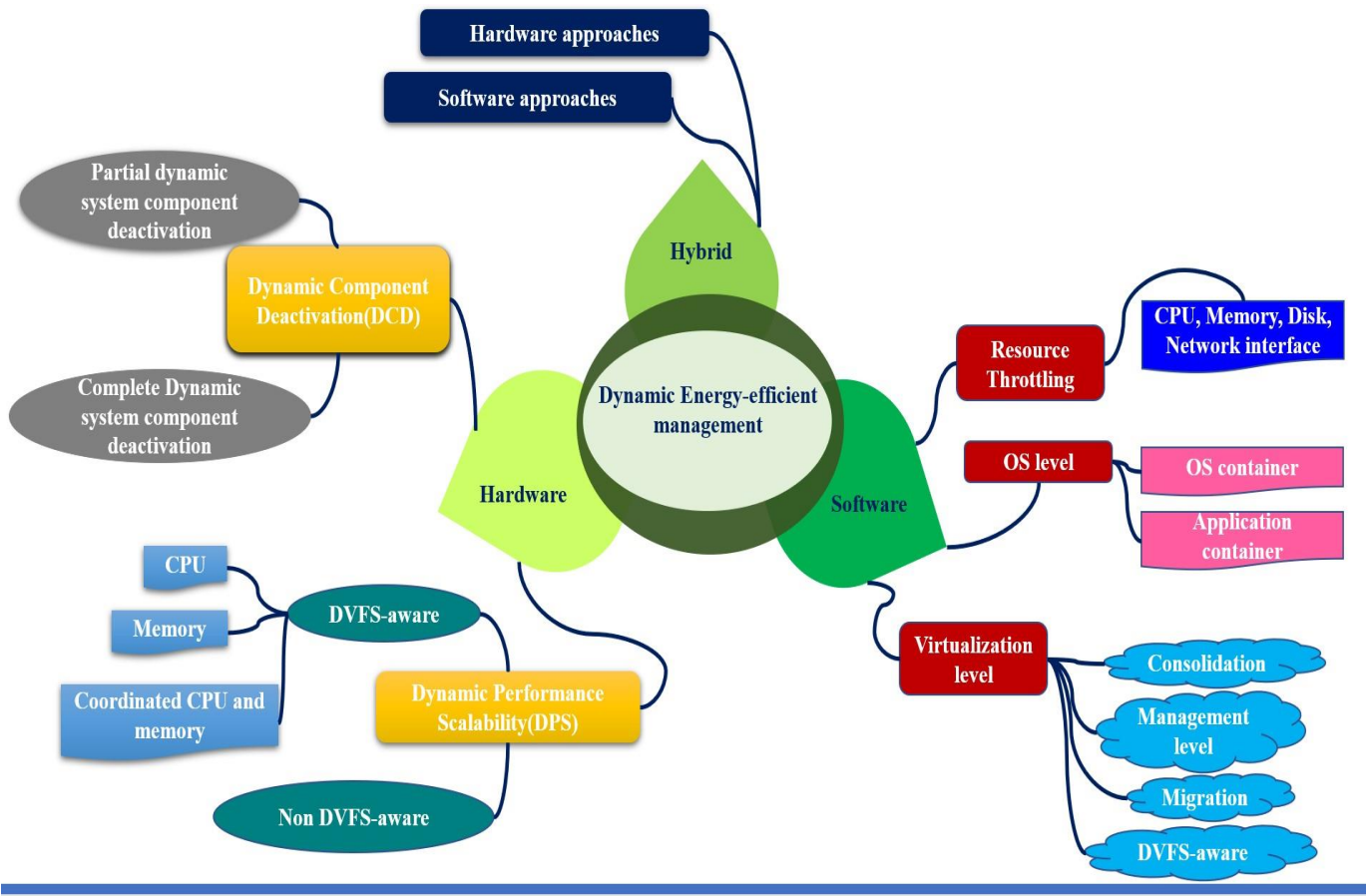

Figure 3: Taxonomy of dynamic resource management solutions in virtualized cloud environments

**Dynamic performance scaling (DPS)**
DPS methods that can be applied in computer components support dynamic adjustment of their performance rationally based on the power consumption and the requested resources [7], [43]. Instead of complete deactivations, just the clock frequency of some components, such as the CPU, is allowed to be decreased or increased along with adjustments of the supply voltage in cases when the resources are not fully utilized. DPS is grouped into both Dynamic Voltage and Frequency Scaling (DVFS) aware techniques and non-DVFS aware techniques.

DVFS is an effective technique at the system level used for both CPU and memory [4]. This method enables dynamic power management by changing the supply voltage or the processor's operation frequencies and memory. Hence, in our proposed taxonomy, we address DVFS at three levels: DVFS of CPU, DVFS of memory, and DVFS of coordinated CPU and memory.

1. **Dynamic voltage and frequency scaling of CPU**
   In this technique, the CPU's voltage and frequency are setting up dynamically proportional to the workload [4], [45]. Thus, the CPU frequency is adjusted to help lessen overall energy consumption for executing the tasks until the user's deadline. Ebbing overall energy consumption is the primary purpose of the algorithms. It is crucial to heed that the DVFS technique is not always energy efficient because scaling the CPU frequency might increase execution and CPU idle time.

   2. **Dynamic voltage and frequency scaling of memory**
      In addition to the CPU, the memory consumes considerable energy [19, 54]. In this technique, the system's memory speed is well adjusted according to the memory-intensive workload's peak power consumption. Employing this technique for workloads that are not memory-intensive would lead to performance degradation as well as higher energy consumptions.

   3. **Dynamic voltage and frequency scaling of coordinated CPU and memory** Coscale has been introduced by [46] for the first time; it jointly applies DVFS on both CPU and memory concurrently, intending to diminish overall energy consumption. The frequency of each core and memory bus is selected to increase the saving in the overall system's energy. The selected frequencies are not always the lowest ones. One of the most noticeable features of the Coscale can be mentioned as balancing the system and its utilization power. It is done by searching the CPU and memory frequency setting space and fixing the components' voltage according to the picked frequencies.

- **Software level approaches**
  Many DPM algorithms can be performed at the hardware level as a part of the circuit [43]. Nevertheless, it is difficult to implement any modification and reconfiguration at the hardware level. Therefore, there are some strong reasons for migration to software-level solutions. In our taxonomy, the techniques in the software level are categorized into resource throttling, OS container level management, application container level management, and virtualization level management which are described in the following sections.

**Resource throttling**

Resource throttling is beneficial to make cloud computing a great deal more energy-efficient. It controls how users are permitted to consume cloud resources in various ways at either hardware or software levels to meet the performance requirements whilst reducing the energy consumption [47]. Several parameters can be throttled in a cloud environment, yet we consider the CPU, memory, disk, plus network interface parameters in our proposed taxonomy. To prevent resources from being overwhelmed by access requests, techniques can be applied to throttle the amount of granted resources.

**Operating system (OS) level**

In 1996, Intel, Microsoft, and Toshiba published the first version of the Advanced Configuration and power interface (ACPI)[43]. ACPI is an open standard that defines a unified OS, centric device configuration, and power management interface [48]. It further describes an independent platform interface for monitoring, discovery, power management, plus configuration of hardware [49]. It is designed to permit OSs to configure and control each hardware component, dislocating both the predecessor's plug and play (PnP) energy management and the advanced power management (APM). The primary purpose of ACPI is to enable the whole computing system to implement all of the DPM capabilities and efficiently develop the power-managed system. ACPI provides an interface for software developers to adjust the system's power states. It defines some power states that can be applied at run-time, containing processor operational states (P-states), sleep-states (S-states), global states (G-states), device states (D-states), and processor idle states (C-states). The two of the most remarkable states in DPM pertain to C-state and P-state [48], [49]. The power management methods in our taxonomy at the OS level are sub categorized into OS container and application container that will be described in the following subsections.

   1. **OS container** Containerization is a lightweight technology that virtualizes and manages applications, has recently been successful in cloud environments [50]. Container technology is currently employed to decrease the difficulty of software deployment and portability of applications in cloud computing infrastructure [51]. In [5], they have proposed Container as a service (CaaS) architecture, consisting of physical hosts, virtual machines (VMs), containers, applications, and their workloads. CaaS can be located between IaaS and PaaS layers; while, IaaS provides virtualized compute resources, and PaaS provides application runtime services, CaaS stick these two layers together by providing isolated environments for the deployed applications (or different modules of an application). PaaS has accelerated the development of applications without any requirements to manage underlying infrastructure [4], [40]. Containers can be run on both Physical Machines (PMs) and VMs, and they create an isolation space for the applications and services running on them. Containers are the building blocks of OS-level virtualization that propose an individual virtualization environment, which does not need a monitoring device like a hypervisor. Containerization technology is implemented in large-scale corporations such as Google and Facebook. One of the most significant advantages of containerization compared to virtualization is its software layer compactness, which impacts lighter overhead to the system. Another benefit that can be mentioned for OS containerization is sharing the host kernel.

2. **Application container** The application container is dedicated to one process, while the OS container runs a number of processes and services [4]. Application container is a new technology with some advantages such as lightweight, more straightforward configuration and management, and a significant reduction in startup time. As the workloads are unpredictable, containers are provided with auto-scaling that conduces to diminishing of resource wastage and gradual increase in energy consumption. Hence, designing optimal container placement algorithms is the biggest challenge for cloud providers. Containers can be applied fast owing to their low overhead. In CaaS model, applications are normally executed inside containers while containers are placed in virtual machines [5].

- **Hybrid: Software and Hardware**
  In our proposed taxonomy at DPM level, we have presented a hybrid approach that is a combination of virtualization and OS-level approaches. Hybrid approaches use both hardware and software techniques at the same time. As shown in our comparison tables, some works have concentrated on both techniques; thus, we defined "hybrid" as a new approach in this domain.

## 4 Virtualization level

In this section, we present virtualization taxonomy, as shown in figure 4, Virtualization is an evolving technology in the IT world [52]. In cloud computing, virtualization strategy is creating a virtual (rather than actual) version of a resource or device, such as a server, an operating system, a storage device, or a network. Virtualization technology is one of the key features of efficient resource management in cloud data centers that can improve hardware efficiency through resource sharing, migration, and consolidation [7]. The virtualization technique makes it possible to abstract the operating system and applications running from the underlying hardware. This technology decreases the operational costs by sharing physical resources, including CPU, RAM, and I/O, plus reducing the number of physical machines. We split our taxonomy at the virtualization level into four main subjects: 1) whether the solutions in this level are DVFS aware or not, 2) the management level of the solutions, 3) the considered migration cost and type in the solutions, 4) the consolidation level as well as the considered consolidation sub-problems. The DVFS technique scales both the frequency and voltage of processors according to the computation requirements and reduces the energy dissipations. We consider three sub-levels for the management levels of virtualization, consisting of VM management, Virtual infrastructure (VI) management, and cloud management [53] . VM management layer supplies the required procedures for managing VMs on a single PM. The duties of the VI management layer are scheduling and managing VMs on multiple PMs. The Cloud management layer provides secure and remote interfaces for monitoring, controlling, and creating virtualized resources. We want to emphasize the importance of this division by highlighting that DVFS and consolidation procedures are performed sequentially and separately in the VM management and Cloud management layers, which may have harmful cross-side effects on each other [53] . More precisely, the DVFS governor scales the processor's frequency dynamically in the VM management layer according to the host global CPU load and regardless of the VM local load. The consolidation technique can be applied at the servers, racks, data centers, or VMs levels; furthermore, the remaining nodes can be turned off or put in the low power mode.

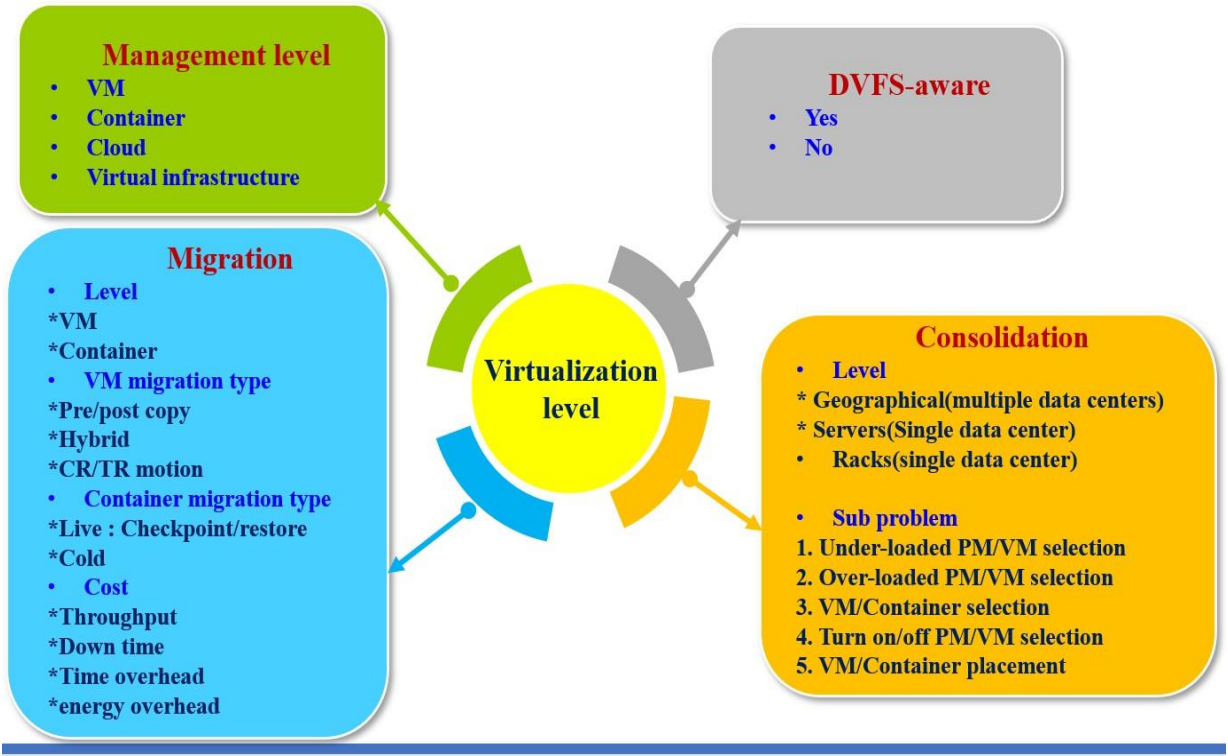

Figure 4: Taxonomy of dynamic energy-efficient resource management in the cloud environment

## 4.1 Migration

The notable characteristic of virtualization that makes it possible to manage resources efficiently can be referred to as both virtual machine and container migration and consolidation [54]. In the consolidation technique using VM migration, all the VMs residing on underutilized PMs are transferred to other active PMs. The idle PMs are then turned off or put in the low-power consumption mode to reduce total energy consumption. Three hardware states are carried through migration, including memory, application, and network interface cards (NIC) [55].In the same way, in container consolidation, the entire containers placed in under-loaded VMs are sent to other VMs owing to turning off the underloaded VMs. The VM migration mechanism can be classified into both live (hot) migration and non-live (cold) migration [55]. In Non-live migration, a VM is paused first at the source PM and resumes at the destination PM after all the required workloads are migrated. Nevertheless, in live migration, the VM resumes its service during the migration. The optimally utilized nodes consume less energy than either over-utilized or under-utilized PMs [56]. Another remarkable profit of VM migration is the possibility of mitigating hot spots in data centers by migrating some loads of over-utilized nodes to other less utilized nodes [55]. More advanced live migration mechanisms do not require a noticeable pause in virtual machines; also, it replies to the customer's requests during the transition [57]. However, in non-live migration, the response time is too long compared with live migration, and there is no high availability. Generally, migration techniques can be used in order to reach different goals, such as green computing, load balancing, fault-tolerant, and real-time server maintenance [58].

Live-migration approaches that we have considered in our taxonomy are including pre-copy, post-copy, hybrid, as well as CR/TR-motion that will be presented in the following subsections.

### 4.1.1 VM Migration type

- **Pre-copy approach**

  Pre-copy is one of the common techniques in live migration mechanism [57]. It is called live owing to the VM does not stop during the migration [59]. First, all of the memory pages are transferred, and then the modified pages are transmitted in each iteration. The pre-copy approach includes warm-up and stop and copy phases[55]. Hypervisor copies just all memory pages from the source node without interrupting the VM, and through this function, the changes on these pages are recorded (in terminology "dirty page"). In the subsequent step, modified pages are sent. The hypervisor suspends the VM in the source node, and in the last iteration, any remaining pages, processor states such as register values are transferred. Finally, the original VM in the source node is discarded, and the VM in the destination node will resume.

- **Post copy approach**

  One of the significant drawbacks of the pre-copy approach can be referred to as the warm-up phase is too long, while the post-copy approach outperforms in this regard [55]. The post-copy approach's migration time is far less than the pre-copy mechanism, yet its downtime is beyond pre-copy inasmuch as each memory page is transferred just once [60]. Contrary to pre-copy, post copy transfers the CPU and device states into the destination node at the first step and commences the destination node's execution in the second step. One of this approach's principal problems is performance degradation when there is a large working set workload. The processes of post copy approach are as follows [57]:

  1. The hypervisor stops the VM at the source node.
  2. At first, the processor states are copied to the destination VM, and this VM starts to work.
  3. If the VM accesses the page that does not exist in the destination node, the page fault occurs, and the post copy mechanism handles it. Ultimately, VM's page is transferred to the destination node (on-demand paging).

- **Hybrid approach**

  This approach is a hybrid of post-copy, and pre-copy methods [55]. Similarly to the pre-copy's first iteration, the whole pages are transferred to the target node in this approach. What is more, this method can avoid performance degradation. Compared with the previous two methods, the performance penalty only after the VM starts on the destination node can be less [55]. The hybrid method accomplishes more memory transition rather than the post-copy one, although it migrates such pages fetched by transporting all memory at the first step. The processes of the hybrid approach are as follows [61]:

  1. Preparation phase.
  2. Bounded pre-copy rounds phase: memory pages send in iterations to the target host. There is a maximum number of iterations until this repetitive transition resumes.
  3. VM state transfer phase: the VM processor state is transferred to the target host.
  4. VM resume phase: this happens at the target host after the VM's processor state is transferred.
  5. On-demand paging phase: this phase is similar to the post copy mechanism. It rises where the VM at the destination host requires transferring the remaining memory pages from the source.

- **CR/TR motion approach**

  CR/TR motion is a novel procedure in the live VM migration category that stands for Check Pointing/Recovery and Trace/Replay [61]. Previous approaches were useful in a local area network (LAN) environment; nonetheless, they would cause a long period of downtime in a vast area network (WAN) environment [62]. This scheme emerged, leading to a fast and transparent transition for both LAN and WAN environments [63]. CR/TR motion can significantly decrease the migration downtime and network bandwidth consumption. It transfers an execution log instead of VM memory and repeats the log on the destination node to generate the same state on the target VM [55]. First, it exploits the memory image of the target VM and sends it to the destination node. Following that, the virtual machine monitor (VMM) level mechanism registers the VM's execution and then frequently transfers the log to the destination. The logging mechanism is based on the log/reply system, which confiscates and replays all non-deterministic events that can affect the VM's execution, such as virtual interrupts. The replay mechanism on the destination repeats the VM execution from the transferred log. CR/TR motion begins the stop and copy phase when the log size is smaller than the predefined size [64].

### 4.1.2 Container Migration type

Let us consider container migration with two general types: live and clod. In the process of cold migration, first, the container stops at the source node, then copies its file system to another node, the destination node. Ultimately, the container at the destination node starts. One of the most notable ways in live migration goes back to checkpoint and restart, which enables it to transmit running container from one physical machine to another without stopping the container. Indeed, the container's file system copies to another server, and the container is restarted on another physical machine from the file [65].

### 4.1.3 Migration cost

The migration process includes some costs that should be considered [66]. Our proposed taxonomy classifies migration costs into four categories: power overhead, migration time overhead, downtime, as well as throughput. The migration technique increases the utilization of the resources (CPU, network interface), contributing to high-energy consumption [67]. The most energy consumption within the migration occurs by the time the VM states are stored and sent to the target host. Since the target host sends acknowledgment and resource availability checking to the source host for migration commencement declaration, the target host consumes more energy. Besides, there is a time span that two different hosts consume energy for the same VM. The cost of VM migration depends on two factors: the power consumed by network devices and the migrated memory size. Generally, the cost of migration is increased for larger RAM sizes and is decreased by raising the network bandwidth [62]. Due to migration operation not being power-free, the energy consumed during migration must be considered once designing or developing energy-saving techniques[66].

## 4.2 Consolidation

The consolidation technique is an act that compacts and combines some units to the integral unit by turning off the underutilized PM or sitting them in the sleep mode. It is formulated as a bin-packing problem [68]. The bin-packing problem is to put several items into a finite number of bins, and the minimal bins are used. In the VM consolidation problem, each VM and host is assumed as an item and a bin, respectively. Virtualization technology is the base of consolidation [54]. This approach maximizes resource utilization and minimizes energy consumption. This survey considers consolidation at three levels including rack, server, and data center (which are geographically distributed).

### 4.2.1 Consolidation level

Consolidation at these levels aims to increase resource utilization and decrease energy consumption. For optimization of the data center's operation, it is essential to consider and minimize all components' energy, including information technology (IT), cooling system, and network equipment [39]. Consolidation of multiple active VMs can be executed on either a single physical server or racks. In rack consolidation, it is also possible to save the cooling system's energy and relevant switches of idle racks by turning them off or putting them in a low-power mode [39], [40].

### 4.2.2 Consolidation sub problems

Virtualization technology makes consolidation of VMs on PMs possible or consolidation of containers on both VMs and PMs. Many researchers have taken advantage of this technique to save the energy consumption in cloud data centers. The methods are categorized according to the subproblems that they explore [4].Simillar to [44], we consider five sub-problems for consolidation in our proposed taxonomy:

1. Determination of underloaded PMs/VMs: if the host is considered as an underloaded host, all the VMs/containers residing on this host/VMs should be migrated from it and then host/VM and it should be put in either the low-power or turn-off mode to save energy.

2. Determination of overloaded PMs/VMs: if the host/VM is considered as an overloaded host/VM, some VMs/containers should be migrated from this host/VM to other active or reactivated hosts/VMs to prevent SLA violation.

3. VM/container selection for migration from either overloaded or underloaded hosts/VMs.

4. Determination of physical nodes that should be put in switched ON or OFF modes.

5. Finding suitable placement for migrating VMs/containers on active or reactive hosts/VMs.

# 5 Comparing the state of the art

In this section, we present a comparison among the state of the arts on energy-efficient cloud resource management solutions taking advantage of our proposed taxonomy presented in previous section. Tables 2 and 3 present our measurements on selected papers published in 2015 through 2021 at the power management level as well as virtualization level, respectively. Table 2 reveals that the selected papers are compared with each other regarding their considered goals, dynamism, and granted workload and resource types. Table 3 analyzes the selected papers with each other respecting being DVFS aware, as well as their considered management level, migration types, plus consolidation solution. used the acronyms shown in tables 2, 3, and 4 for the considered objects, the migration costs, and considered consolidation sub problems, respectively. Besides, we utilize NM (Not Mentioned) notation in Tables 5 and 6, where the information is not provided in the reviewed paper.

Table 2: The notations used for resource management goals

| Acronym | Goal |
|---------|------|
| MEC | Minimize Enegy Consumption |
| MPL | Minimize Performance Loss |
| MBP | Maximize Business Profit |
| LB | Load Balancing |

Table 3: The notations used for migration costs

| Acronym | Migration Cost |
|---------|----------------|
| MTO | Migration Time Overhead |
| DT | Down Time |
| EO | Energy Overhead |
| TH | Throughput |

The authors [69] have proposed an energy-aware combinatorial auction-based model for the resource allocation problem in clouds. They have regarded both MEC and MBP as their goals. Besides, at the dynamism level, they have considered virtualization techniques in the software category of DPM. CPU, RAM, Disk, as well as network are the active resources that they have applied in their proposed heuristic methods. The management level in [69] can be referred to as VM, VI, and cloud. Also, they have considered two distributed datacenters and solved VP and VS problems.

Table 4: The notations used for consolidation sub problems

| Acronym | Consolidation sub-problem |
|---------|---------------------------|
| OPS/OVS | Overloaded PM/VM Selection |
| UPS/UVS | Underloaded PM/VM Selection |
| VS/CS | VM/container Selection |
| TPS/TVS | Turn on/off PM/VM Selection |
| VP/CP | VM/container Placement |

The authors [53] have focused on MPC and MPL as their goals. At the dynamism level, they have used DPM techniques, including DPS and virtualization. Also, they have applied the arbitrary types of workload. They further have considered CPU, RAM, and network bandwidth as cloud resources. Their approach is DVFS aware, and their considered management levels are VM, VI, and cloud. The type of migration that they have used is pre-copy. Moreover, they have solved consolidation subproblems OPS, UPS, and VP at the server level.

The authors [70] have considered MPC and MPL as their goals, as shown in table 2. At the dynamism level, they have adopted the virtualization method and used arbitrary workload. Plus, they have applied active resources such as CPU, RAM, Disk, and network. They have only focused on the cloud management level. For the migration types, they have regarded both MTO and EO as migration costs. They have further utilized the pre-copy migration type and solved consolidation subproblems OPS and VP at the server level.

As can be seen in table 5, the authors [71] have measured MPC and MPL as their goals. At the dynamism level, they have applied the virtualization technique. Additionally, they have applied an arbitrary workload type. In the category of active resources, they have considered CPU, RAM as well as network bandwidth. As further demonstrated, their management level is the cloud level. They have considered both downtime and throughput in their proposed work as migration costs. Furthermore, they have implemented the pre-copy approach and did consolidation subproblems OPS and VS at the server level.

The goals that reflected [72] pertains to MPC and MPL. At the dynamism level, the researchers have applied the virtualization technique via arbitrary workload type. They have used CPU, RAM, and network as active resources in their work in addition to the cooling system as a passive resource. What is more, their management level is in the cloud level. They have merely concentrated on migration time overhead as the migration cost. Moreover, they have employed the pre-copy approach for migration and have solved Vms as one of the consolidation subproblem VS at the server level.

MPC and MPL are considered [73]. The researchers have practiced the virtualization technique at the dynamism level. Moreover, they have used arbitrary workload and focused on CPU, RAM, and network as their resources. They further applied their proposed approach at the cloud level and used the pre-copy migration approach, considering downtime overhead as migration cost. Moreover, they have Addressed OPS,UPS, VMs, and VMP consolidation subproblems at the server level.

The authors [41] have just contemplated MPL object. They have implemented both DPS and virtualization procedures at the dynamism level via arbitrary workload types. Besides, they have considered CPU and RAM as resource types. They have used the DVFS-aware technique at the cloud management level. They more have regarded energy overhead as migration cost, yet they have not mentioned the type of migration. The VMP Consolidation subproblem is done in this paper.

Based on our study, the scholars [74] have paid attention to MPC and MPL goals. They have implemented the virtualization technique through an arbitrary workload to verify it at the cloud management level. Besides, they considered active resources containing CPU and RAM. As further demonstrated, they have confronted with OPS, UPS, and VMP consolidation subproblem at server level.

MEC and LB are reflected in [75]. They have come across a solution at the dynamism level by taking advantage of virtualization approach. The authors have not specified their utilized workload; conversely, the active resources they have adopted are CPU, RAM, and network. They have employed their solution not only at VM but also at cloud management level. The type of migration that they have managed pertains to pre-copy, respecting energy overhead as its cost. Another notable part that can be mentioned in this paper is working on VMP consolidation subproblem at the server level.

The goals that are analyzed in [76] goes back to MEC and MPL. Regarding dynamism level, the researchers in this paper deployed the virtualization procedure using the arbitrary workload. Plus, they only focused on active resources such as CPU and RAM. At the migration level, they have applied the pre-copy approach in the cloud management level by considering MTO as cost. In addition, both VMS and VMP consolidation subproblems are addressed in this research work.

MPC and MPL are noted [77]. Similar to previous works, they made use of the virtualization technique utilizing an arbitrary workload prototype. Active resources employed in this research can be referred to as CPU, RAM as well as network. Both VM and cloud management levels are tackled with in this paper. Although they have not mentioned the type of migration, they have considered energy overhead and migration time overhead as its costs. Researchers have also worked on OPS, UPS, and VMP consolidation subproblems at the server level.

The scholars [78] have weighed MEC in addition to MPC as their goal. The approaches at the dynamism level applied by them pertain to DPS and virtualization at both VM and cloud management level. Unlike others, they have utilized batch workload type, and have adopted CPU, RAM, and Disk as the active resources in their work. Moreover, the technique addressed in this paper goes back to DVFS-aware. Also, VMP Consolidation subproblem is addressed at the server level.

The authors [79] have just concerned on MEC goal. Like others, they have used the virtualization technique at the dynamism level through the arbitary workload. CPU and RAM can be mentioned as the active resources that they have deployed. Besides, they have examined their proposed method not only on the VM but also on the cloud management level. Ultimately, they have provided solution for solely the VMP consolidation subproblem.

The researchers [80] have contemplated MEC, MBP, as wll as LB goals. At the dynamism level, they have applied the virtualization procedure. Also, they made use of arbitrary workload, yet they have not discussed their utilized resource types. Their management levels are VI and cloud.In contrast to others, their considered consolidation level is geographical.

Like most previous works, [81] have focused on MEC and MPL goals. The researchers in this paper have applied the virtualization technique at the dynamism level via arbitrary workload. The only active resource that they have used pertains to CPU. Furthermore, the cloud is their management level. They have considered pre-copy type migration and both time and energy overhead for the migration cost. They have solved OPS, VMS, and VMP consolidation sub-problems.

According to our review, [82] investigated MEC and MPL goals. They have more implemented the virtualization technique at the dynamism level through the arbitrary workload. Similar to most of the earlier works, they have regarded CPU as the active resource type, and their management level is the cloud level. As further identified, they have not mentioned the migration type, yet they have reflected energy overhead as migration cost. Another significant point addressed in this paper goes back to solving TPS and VMP sub-problem at the server level.

Three goals are explored [29], including MEC, MPL, and LB. Moreover, they have applied both DPS and virtualization methods within the arbitrary workload at the dynamism level. The authors have investigated both active and passive resources in their work containing CPU, RAM as an active and cooling system as a passive type. Their proposed approach is DVFS-aware at the cloud management level. Moreover, they have examined OPS and VMP consolidation subproblems

The authors [83] have studied solely MEC goal in their proposed method. Plus, they have employed the DVFS-aware DPS procedure. Unlike the rest of the papers, this work used real-time workload. Besides, they have applied their proposed approach at both VI and cloud levels.

As we deduced from [84], solely the MPE is regarded in this paper. Plus, they have applied the DVFS technique at the dynamism level, a subset of DPS in our taxonomy. The workload type that they have applied pertains to HPC. They have further utilized both CPU and RAM as active resources.

The scholars [85] have reflected MEC goal. The approach employed in their proposed work goes back to virtualization at the dynamism level, including server consolidation via real-time workload. Another noticeable point in their work is that they have considered not only CPU as an active resource but also RAM. Besides, they have confronted the TPS consolidation.

The MBP and LB goals are probed [86]. The authors have also employed a virtualization approach at the dynamism level, including workload consolidation at geographically distributed data centers through an arbitrary workload.

The goals that are chosen [87] can be referred to as MEC and MPL. Moreover, they have applied the virtualization technique at the dynamism level in the software approach's subset and validated their proposed technique through the arbitrary workload. Unlike the most previous works, they are cared about passive resources such as the cooling system and power supply unit. They have also handled their proposed approach at three management levels consisting of VM, VI as well as the cloud. From the cost point of view, solely an energy overhead was noticed by the authors. As further concluded, the VMP consolidation subproblem is studied by them at the geographically distributed data centers.

MEC and MBP goals are remarked by researchers [88]. The authors explored the DVFS aware virtualization method within the arbitrary workload in their proposed approach. They have paid attention to both active resources, CPU and RAM, and passive resources, such as cooling system and co-location interference.In addition, they have confronted with the VMP consolidation subproblem at the geographically distributed data centers.

The scholars [89] are concerned about MEC in addition to MPL goals. They have applied a virtualization approach via arbitrary workload. The active resources operated in their testbed can be mentioned as CPU, RAM, and storage. Three management levels are noticed by them, consisting of VM, VI, as well as the cloud. Another significant point in their proposed work is that they studied VMP consolidation sub problem in geographically distributed data centers.
The authors [90] have addressed MEC and LB goals. They have also implemented virtualization techniques by the arbitrary workload. They have further focused solely on active resources, including CPU, RAM, and storage.

The authors [91] have focused on both MEC and MPL goals. Besides, their proposed approach is at the virtualization level, a subset of software category in dynamic energy-efficient management. Indeed, they have considered joint VM and container consolidation and verified it by the arbitrary workload in the CloudSim simulator. Their considered type migration was pre-copy, plus energy, as well as time overhead, as their migration cost. As further investigated, they have only concentrated on an active resource, CPU. What is more, Cloud and VM are their management level, and the VMs subproblem is solved in this paper at the server level.

From the goals perspective, MEC and MPL are remarked by [92]. They proposed Brownout, which dynamically deactivates or activates optional microservices or containers to handle over- loads and reduce power consumption within a datacenter which belongs to the software level. As more explored, the authors have examined container solely in their proposed approach. Besides, they have chosen the cloud management level. The OPS and TPS consolidation subproblems that are fulfilled in this paper.

The MPL goal is noted [93]. The term of the cost that is estimated in this paper is based on VM's provisioning from the request moment until deprovisioning request. Nonetheless, for static nodes, the cost is measured according to the workload's total scheduling time. They solely contemplated container management in their proposed technique through the Kubernetes platform via batch workload. The active resources that they have applied in their work can be referred to as memory as well as CPU.

Another study that have met CaaS environment goes back to [94] that measured the MEC and LB goals . The authors have proposed a renewable energy-aware multi-indexed job classification and scheduling scheme using CaaS for data centers. Thus, they have addressed CP and OPS as a subset of the consolidation problem. Plus, they have verified their proposed scheme utilizing extensive simulations for 3 geo-distributed data centers having 200 heterogeneous servers via Google workload traces. They have noted both passive and active resources including cooling system, CPU, and network.

As we perceived from [95] the MPL goal including, improving the utilization rate of resources and enhancing the user experience are the targets that the author covered. They have proposed a live container migration algorithm named Gray-Markov prediction model to foreseen the probability that dirty pages are modified again.Further, the type of migration is pre-copy, and lessening both downtime, iteration as well as total copy time are considered as migration costs in their method.

The authors [96], have focused on consolidation technique in containerized datacenters. They have investigated a new algorithm for container allocation concerning both MEC and MPL goals. Besides, they have proposed a mathematical model which estimates the energy consumption of containers executed in VMs. The proposed migration technique is performed only if the migration cost capable of recovering to save energy. They have further applied CPU and RAM as active resources. The consolidation subproblem that they have considered is CP with considering of EO as migration cost at the server level.

The goals which are addressed [97] go back to both MEC and MBP. The main contribution of this work is that they have developed an experimental setup to execute and assess the performance of the Linux Container Hypervisor (LXD) and checkpoint/restore (CR) container based migration methods and comparing it with a pre-copy VM migration scheme. Their applied workload is real-time while considering CPU and RAM as active resources in their proposed technique. They have adopted LXD/CR mechanism for container migration. The CP subproblem has addressed by [97] at the server level.

The goal that is targeted [98] can be referred to as MPL. The authors of this paper have applied the virtualization method at the OS level. Further, they considered both CPU and memory as actives resources to evaluate their proposed approach at container level management. They have concerned about DT as well as TH cost in CR/TR migration. Plus, they have considered the CP subproblem at the geographically distributed data centers.

Table 5: Comparing the state of the arts in the power management level

| Ref | Year | Goals | | | | Dynamism | | | | | | | | | | Workload | | | | Resources | | | | | | |
|---|---|---|---|---|---|---|---|---|---|---|---|---|---|---|---|---|---|---|---|---|---|---|---|---|---|---|
| | | | | | | DPM | | | | | SPM | | | | | | | | | Active | | | | Passive | | |
| | | MEC | MPL | MBP | LB | Hardware | | Software | | | Hardware | | | Software | Arbitrary | HPC | Batch | Real-ime | CPU | RAM | Disk | Network | Cooling system | PSU | Others |
| | | | | | | DCD | DPS | OS level | Resource throttling | Virtualization | Circuit | Logic | Architecture | | | | | | | | | | | | | |
| [69] | 2021 | ✓ | | ✓ | | | | | | ✓ | | | | | ✓ | | | | ✓ | ✓ | ✓ | ✓ | | | |
| [53] | 2017 | ✓ | ✓ | | | | ✓ | | | ✓ | | | | | ✓ | | | | ✓ | ✓ | | ✓ | | | |
| [70] | 2015 | ✓ | ✓ | | | | | | | ✓ | | | | | ✓ | | | | ✓ | ✓ | ✓ | ✓ | | | |
| [71] | 2016 | ✓ | ✓ | | | | | | | ✓ | | | | | ✓ | | | | ✓ | ✓ | | ✓ | | | |
| [72] | 2016 | ✓ | ✓ | | | | | | | ✓ | | | | | ✓ | | | | ✓ | ✓ | | ✓ | ✓ | | |
| [73] | 2018 | ✓ | ✓ | | | | | | | ✓ | | | | | ✓ | | | | ✓ | ✓ | | ✓ | | | |
| [41] | 2017 | | ✓ | | | | ✓ | | | ✓ | | | | | ✓ | | | | ✓ | ✓ | | | | | |
| [74] | 2018 | ✓ | ✓ | | | | | | | ✓ | | | | | ✓ | | | | ✓ | ✓ | | | | | |
| [75] | 2018 | ✓ | | | ✓ | | | | | ✓ | | | | | ✓ | | | | ✓ | ✓ | | ✓ | | | |
| [76] | 2018 | ✓ | ✓ | | | | | | | ✓ | | | | | ✓ | | | | ✓ | ✓ | | | | | |
| [77] | 2018 | ✓ | ✓ | | | | | | | ✓ | | | | | ✓ | | | | ✓ | ✓ | | ✓ | | | |
| [78] | 2017 | ✓ | ✓ | | | | ✓ | | | ✓ | | | | | | | ✓ | | ✓ | ✓ | ✓ | | | | |
| [78] | 2017 | ✓ | | | | | | | | ✓ | | | | | ✓ | | | | ✓ | ✓ | | | | | |
| [80] | 2017 | ✓ | | ✓ | ✓ | | | | | ✓ | | | | | ✓ | | | | | | | | | | |
| [81] | 2018 | ✓ | ✓ | | | | | | | ✓ | | | | | ✓ | | | | ✓ | | | | | | |
| [82] | 2018 | ✓ | ✓ | | | | | | | ✓ | | | | | ✓ | | | | ✓ | | | | | | |
| [99] | 2018 | ✓ | ✓ | | ✓ | | | | | ✓ | | | | | ✓ | | | | ✓ | ✓ | | | ✓ | | |
| [83] | 2018 | ✓ | | | | | ✓ | | | ✓ | | | | | | | | ✓ | ✓ | | | | | | |
| [84] | 2018 | ✓ | | ✓ | | | | | ✓ | | | | | | | ✓ | | | ✓ | ✓ | | | | | |
| [85] | 2017 | ✓ | | | | | | | | ✓ | | | | | | | | ✓ | ✓ | ✓ | | | | | |
| [86] | 2016 | | | ✓ | ✓ | | | | | ✓ | | | | | ✓ | | | | | | | | | | |
| [89] | 2017 | ✓ | ✓ | | | | | | | ✓ | | | | | ✓ | | | | | | | | ✓ | ✓ | ✓ |
| [88] | 2018 | ✓ | | ✓ | | | | | | ✓ | | | | | ✓ | | | | ✓ | ✓ | | | ✓ | | ✓ |
| [79] | 2017 | ✓ | ✓ | | | | | | | ✓ | | | | | ✓ | | | | ✓ | ✓ | ✓ | | | | |
| [90] | 2017 | ✓ | | | ✓ | | | | | ✓ | | | | | ✓ | | | | ✓ | ✓ | ✓ | | | | |
| [91] | 2020 | ✓ | ✓ | | | | | | | ✓ | | | | | ✓ | | | | ✓ | | | | | | |
| [92] | 2019 | ✓ | ✓ | | | | | | | ✓ | | | | | ✓ | | | | | | | | | | |
| [93] | 2020 | ✓ | ✓ | | | | | | | ✓ | | | | | | | ✓ | | ✓ | ✓ | | | | | |
| [94] | 2018 | ✓ | | | ✓ | | | | | ✓ | | | | | ✓ | | | | ✓ | | | | | | |
| [95] | 2017 | | | | | | | | | ✓ | | | | | ✓ | | | | | ✓ | | | | | |
| [96] | 2019 | ✓ | ✓ | | | | | | | ✓ | | | | | ✓ | | | | ✓ | ✓ | | | | | |
| [97] | 2019 | ✓ | | ✓ | | | | ✓ | | ✓ | | | | | | | | ✓ | ✓ | ✓ | | | | | |
| [98] | 2020 | | ✓ | | ✓ | | | ✓ | | ✓ | | | | | ✓ | | | | ✓ | ✓ | ✓ | | | | |

Table 6: Comparing the state of the art in the virtualization level

| Ref | Year | DVFS-aware | Management Level | | | | Migration | | | | | | Consolidation | |
|---|---|---|---|---|---|---|---|---|---|---|---|---|---|---|
| | | | Container | VM | VI | Cloud | Considered Cost | | | | Technique | Type | Level | Subproblem |
| | | | | | | | MTO | DT | EO | TH | | | | |
| [69] | 2021 | | | ✓ | | ✓ | | | | | VM | | Geographical | VP-VS |
| [53] | 2017 | ✓ | | ✓ | ✓ | ✓ | | | | | Pre-copy | VM | server | OPS-UPS-VP |
| [70] | 2015 | | | | | ✓ | ✓ | | ✓ | | Pre-copy | VM | server | UPS-VP |
| [71] | 2016 | | | | | ✓ | | ✓ | | ✓ | Pre-copy | VM | server | OPS-VS |
| [72] | 2016 | | | | | ✓ | ✓ | | | | Pre-copy | VM | server | VS |
| [73] | 2018 | | | | | ✓ | | ✓ | | | Pre-copy | VM | server | OPS-UPS-VS-VP |
| [41] | 2017 | ✓ | | | | ✓ | | | | ✓ | NM | VM | Server | VP |
| [74] | 2018 | | | | | ✓ | | | | | NM | VM | server | OPS-UPS-VP |
| [75] | 2018 | | | ✓ | | ✓ | | | ✓ | | Pre-copy | VM | server | VP |
| [76] | 2018 | | | | | ✓ | ✓ | | | | Pre-copy | VM | server | VS-VP |
| [77] | 2018 | | | ✓ | | ✓ | ✓ | | ✓ | | NM | VM | server | OPS-UPS-VP |
| [78] | 2017 | ✓ | | ✓ | | ✓ | | | | | NM | VM | Server | VP |
| [80] | 2017 | | | ✓ | | ✓ | | | | | NM | VM | Server | VP |
| [81] | 2018 | | | | ✓ | ✓ | | | | | NM | VM | Geographical | NM |
| [82] | 2018 | | | | | ✓ | ✓ | | ✓ | | Pre-copy | VM | Server | OPS-VS-VP |
| [99] | 2018 | | | | | ✓ | | | ✓ | | Pre-copy | VM | server | TPS-VP |
| [83] | 2018 | ✓ | | | | ✓ | | | | | NM | VM | Server | OPS-VP |
| [84] | 2018 | ✓ | | | ✓ | ✓ | | | | | | | | |
| [85] | 2017 | ✓ | | | | ✓ | ✓ | ✓ | | | Pre-copy | VM | Server | VS |
| [86] | 2016 | | | | | ✓ | | | | | | VM | Server | TPS |
| [89] | 2017 | | | | | ✓ | | | | | NM | VM | Geographical | VP |
| [88] | 2018 | | | ✓ | ✓ | ✓ | | | ✓ | | NM | VM | Geographical | VP |
| [79] | 2017 | | | ✓ | ✓ | | | | | | NM | VM | Geographical | VP |
| [90] | 2017 | | | ✓ | | ✓ | | | | | NM | VM | Geographical | VP |
| [91] | 2020 | | ✓ | ✓ | | ✓ | ✓ | | ✓ | | Pre-copy | VM,Container | Server | VS-CP |
| [92] | 2019 | | ✓ | ✓ | | | | | | | NM | Container | Server | OPM-CS |
| [93] | 2020 | | ✓ | ✓ | | | | | | | Pre-copy | Container | Server | CP |
| [94] | 2018 | ✓ | ✓ | | ✓ | ✓ | ✓ | | | | Pre-copy | Container | Geographical | OPS-CP |
| [95] | 2017 | | ✓ | ✓ | | | | ✓ | | | Pre-copy | Container | Server | CS |
| [96] | 2019 | | ✓ | ✓ | | | | | ✓ | | Pre-copy | Container | Server | CP |
| [97] | 2019 | | ✓ | ✓ | | ✓ | ✓ | ✓ | | | Pre-copy, CR | Container, VM | Server | CP |
| [98] | 2020 | | ✓ | | | | | ✓ | | ✓ | CR/TR | Container | Geographical | CP |

It can be inferred from Table 5 that not only minimizing power consumption but also the performance loss were the main aims reflected in the state of the arts for resource management in cloud environments. Nonetheless, few papers consisting of, [99], [80], [75], as well as [86], [98],[94], and [90] regard the load-balancing goal along with declining power consumption and performance loss. Also, [69], [80], [84], [86], [88], and [97] counted maximizing business profit. As further demonstrated, one of the most remarkable points goes back to the researcher's attention to DPM techniques as an offspring of its superiority compared with SPM techniques. What is more, only [98] and [97] utilized resource management techniques at the OS level. further, none of the papers employed hybrid simultaneous hardware and software DPM techniques except [53]. Another significant case that can be mentioned as considering only arbitrary workload type by most of the researchers in their work. Indeed, they neglected batch, real-time, as well as HPC workload types except [78] and [93] papers that noticed batch workload type, plus [83], [85] and [97] papers that marked real-time workload. Ultimately, solely the authors in [84] validated their proposed approach through HPC workload type. Moreover, it can be inferred from Table 5 that most of the recently selected papers only analyze the power consumed over active resources such as CPU, memory, network, and disk storage. While, most of the state of the arts regarded the power consumed in passive peripherals. The researchers in [72], [99], [89], and [88] considered the power consumed in the cooling systems. To add, [89] considered only passive resources, including cooling systems and PSU , and [88] focused on both the cooling system and other passive peripherals.

As can be seen in Table 6, a few papers are addressed DVFS-aware approach in their proposed cloud resource management solutions. Further, most of the articles introduced resource management solutions only at the cloud level without considering either VM or VI level. By contrast, just [53] and [88] regarded all three management levels, including VM, VI, as well as cloud. It can also be deduced from Table 6 that different migration costs are considered in the selected papers. As more displayed, all the studied articles applied the pre-copy migration procedure in their solutions except [97] and [98] applied CR/TR approach. Additionally, [81], [90], [94], [79], [88], [89], [98], and [69] addressed consolidation technique at the geographical level, yet the rest of the papers concentrated only on the server level. As further explored, [91],[92], [93],[94], [95],[96],[97], and [98] remarked container level management along with container consolidation in their work; among them merely [91] applied joint VM and container migration approach, but others employed VM and container migration separately.

# 6 Future scope and Conclusion

This section presents some research issues and challenges concerning energy-efficient resource management methods in the cloud environments. Although notable progress has been accomplished in applying containerization to the cloud computing systems and the adaptive management of resources and applications is widely developed; there are still many research gaps and challenges in this area needed to be further investigated as discusses below.

- **Multiple system resources**: Due to the broad admission of multi-core CPUs, developing energy-efficient resource management approaches plays a crucial role in leveraging such archi- tectures. To optimize a data center's operation, it is critical to reflect and lessen all energy elements consumption, including the cooling system and power supply units, as passive re- sources. To add, RAM, disk storage, and network equipment as active resources are usually overlooked by researchers. From the perspective of active resources, current works mostly focus on CPU. More resource types, like memory, network, storage, and GPU need to be regarded as parameters to create more comprehensive resource management.

- **Rack consolidation and geographically distributed data center**: Many big data analysis applications involve analyzing a large volume of data generated in a geographical-distributed data center. Besides, plenty of data-intensive applications, such as social networks, involve large data sets in multiple geographically distributed cloud data centers. As a case in point, Facebook receives terabytes of text, image, and video data every day from users worldwide. Another noticeable future research direction goes back to the exploration of cloud environment geographically distributed data centers and rack consolidation in addition to server and VM consolidation to make it possible to provide more reliable services in greener data centers.

- **System workload**: Most of the current papers applied arbitrary workloads in their study; conversely, this is a crucial issue to consider other workload types in addition to arbitrary workloads containing the batch, HPC, plus real-time application workloads.

- **Security and privacy**: Over the years, the ever-increasing growth of cloud data centers utilized by famous corporations such as Google, Facebook, and Microsoft, can result in rise of new different administrative and security. So, addressing the security concerns which are become more and more complicated by development of new containerized services, such as Distributed Denial of Service (DDoS), has become an important issue to be considered in future research directions.

- **Cognitive approach contributing to Joint VM and container consolidation**: Container consolidation is an evolving technology which has a great deal better performance than VM consolidation in the light of energy consumption and performance loss. Besides, the research in [91] has justified that joint VM and container consolidation outperforms individual VM or container consolidation approaches, regarding energy consumption and QoS. By contrast, applying artificial intelligence, or machine learning, to make a cognitive decision for simultaneous migration of both VM and container is a hot research topic.

- **Container security**: There are some levels for container ecosystem security including image, registry, orchestrator, container and host OS. For instance, container technologies like Docker and Kubernetes accelerate the development and deployment of application; hence, their security issues play a notable role in software development and cloud industries. The research in this field can be directed in two levels including protecting a container from the security attacks of its applications and protecting a physical server from the security attacks of its containers.

## 6.1 Conclusion

In this chapter, we proposed a holistic taxonomy for cloud resource management and then we surveyed 32 recent articles in the literature related to energy-efficient resource management procedures based on the proposed taxonomy. Since the research has dramatically advanced in the field of resource management techniques in cloud computing, there is a demand for a systematic review to evaluate, update and combine the existing literature. On the one hand, the requests to access cloud services are ever-increasing. On the other hand, cloud services are hosted on massive data centers consuming enormous electrical power. Thereby, efficient energy-aware resource management has become a matter of great concern in cloud environments from single server to data centers and Clouds. This chapter centered on the energy-aware resource management problem in cloud environments and proposed a novel taxonomy and solutions classification. More precisely, first we proposed a holistic taxonomy for energy aware resource management in cloud environment. Then, we surveyed 32 articles in the literature related to energy-efficient resource management and categorized them according to their proposed solution based on a scientific comparison. Eventually, we opened up new research directions by gap analysis and suggesting major drawbacks in the current literature on energy-aware cloud resource management. This chapter rose up to four research questions and answered them, including: 1) The definition of energy-aware cloud resource management. 2) The taxonomy of power-aware resource management solutions. 3) The parameters considered in the literature for energy-aware resource management in cloud environments. 4) The open research directions for energy-aware cloud resource management.
This chapter appears to yield several innovations. One of the most remarkable novelties pertains to presenting a scientific taxonomic survey of recent literature over seven years, 2015 to 2021. This chapter categorized recent research progressions across four main categories: dynamism, workload types, resources, as well as goals. In the light of comparison, we classified articles according to their suggested solution for energy-aware cloud resource management. Moreover, the workload part characterizes literature regarding the considered workload types in the proposed solution. Considered resources revealed the type and variety of taken into consideration resources in the literature. Ultimately, the goal level categorized the literature regarding their main objectives. Further, the dynamism level itself was divided into static and dynamic power management levels. Three subcategories were chosen for dynamic solutions such as hardware, software, plus hybrid levels. Another item belongs to software, which was divided into three subcategories: operating system, virtualization, as well as resource throttling levels. DVFS-aware, management level, migration type and technique, and consolidation were the subcategories of virtualization. Another noticeable novelty can be referred to as regarding container migration in addition to VM migration in the determined period. To add, this chapter categorized the selected papers according to the proposed taxonomy and provided a scientific comparison. Finally, it opened up new research directions by gap analysis and suggesting major drawbacks in the current literature on energy-aware cloud resource management.

**Acknowledgment**
We thank Professor Rodrigo N Calheiros and Kiarash Geraili for their suggestions on improving the manuscript.

# References

[1] M. L. Badger, T. Grance, R. Patt-Corner, and J. M. Voas, *Cloud computing synopsis and recommendations*. National Institute of Standards & Technology, 2012.

[2] P. Mell, T. Grance, *et al.*, "The nist definition of cloud computing," 2011.

[3] A. Khosravi and R. Buyya, "Energy and carbon footprint-aware management of geo-distributed cloud data centers: A taxonomy, state of the art, and future directions," in *Sustainable Development: Concepts, Methodologies, Tools, and Applications*, IGI Global, 2018, pp. 1456–1475.

[4] S. F. Piraghaj, "Energy-efficient management of resources in enterprise and container-based clouds," *PhD, University of Melbourne, Melbourne, Australia*, 2016.

[5] S. F. Piraghaj, A. V. Dastjerdi, R. N. Calheiros, and R. Buyya, "Containercloudsim: An environment for modeling and simulation of containers in cloud data centers," *Software: Practice and Experience*, vol. 47, no. 4, pp. 505–521, 2017.

[6] X. You, Y. Li, M. Zheng, C. Zhu, and L. Yu, "A survey and taxonomy of energy efficiency relevant surveys in cloud-related environments," *IEEE Access*, vol. 5, pp. 14 066–14 078, 2017.

[7] A. Beloglazov, R. Buyya, Y. C. Lee, and A. Zomaya, "A taxonomy and survey of energy-efficient data centers and cloud computing systems," in *Advances in computers*, vol. 82, Elsevier, 2011, pp. 47–111.

[8] R. Buyya, S. N. Srirama, G. Casale, R. Calheiros, Y. Simmhan, B. Varghese, E. Gelenbe, B. Javadi, L. M. Vaquero, M. A. Netto, *et al.*, "A manifesto for future generation cloud computing: Research directions for the next decade," *ACM computing surveys (CSUR)*, vol. 51, no. 5, pp. 1–38, 2018.

[9] R. Yadav, W. Zhang, K. Li, C. Liu, and A. A. Laghari, "Managing overloaded hosts for energy-efficiency in cloud data centers," *Cluster Computing*, pp. 1–15, 2021.

[10] D. Merkel, "Docker: Lightweight linux containers for consistent development and deployment," *Linux journal*, vol. 2014, no. 239, p. 2, 2014.

[11] N. Gholipour, N. Shoeibi, and E. Arianyan, "An energy-aware dynamic resource management technique using deep q-learning algorithm and joint vm and container consolidation approach for green computing in cloud data centers," in *International Symposium on Distributed Computing and Artificial Intelligence*, Springer, 2020, pp. 227–233.

[12] S. Sultan, I. Ahmad, and T. Dimitriou, "Container security: Issues, challenges, and the road ahead," *IEEE Access*, vol. 7, pp. 52 976–52 996, 2019.

[13] A. Bhardwaj and C. R. Krishna, "Virtualization in cloud computing: Moving from hypervisor to containerization—a survey," *Arabian Journal for Science and Engineering*, pp. 1–17, 2021.

[14] E. Casalicchio and S. Iannucci, "The state-of-the-art in container technologies: Application, orchestration and security," *Concurrency and Computation: Practice and Experience*, vol. 32, no. 17, e5668, 2020.

[15] N. Akhter and M. Othman, "Energy aware resource allocation of cloud data center: Review and open issues," *Cluster Computing*, vol. 19, no. 3, pp. 1163–1182, 2016.

[16] S. Singh, A. Swaroop, A. Kumar, *et al.*, "A survey on techniques to achive energy efficiency in cloud computing," in *2016 International conference on computing, communication and automa- tion (ICCCA)*, IEEE, 2016, pp. 1281–1285.

[17] M. Zakarya and L. Gillam, "Energy efficient computing, clusters, grids and clouds: A taxonomy and survey," *Sustainable Computing: Informatics and Systems*, vol. 14, pp. 13–33, 2017.

[18] Q. Shaheen, M. Shiraz, S. Khan, R. Majeed, M. Guizani, N. Khan, and A. M. Aseere, "Towards energy saving in computational clouds: Taxonomy, review, and open challenges," *IEEE Access*, vol. 6, pp. 29 407–29 418, 2018.

[19] S. Puhan, D. Panda, and B. K. Mishra, "Energy efficiency for cloud computing applications: A survey on the recent trends and future scopes," in *2020 International Conference on Computer Science, Engineering and Applications (ICCSEA)*, IEEE, 2020, pp. 1–6.

[20] S. Kaur and S. Bawa, "A review on energy aware vm placement and consolidation techniques," in *2016 International Conference on Inventive Computation Technologies (ICICT)*, IEEE, vol. 3, 2016, pp. 1–7.

[21] M. Zakarya and L. Gillam, "An energy aware cost recovery approach for virtual machine migration," in *International Conference on the Economics of Grids, Clouds, Systems, and Services*, Springer, 2016, pp. 175–190.

[22] D. Hazra, A. Roy, S. Midya, and K. Majumder, "Energy aware task scheduling algorithms in cloud environment: A survey," in *Smart Computing and Informatics*, Springer, 2018, pp. 631–639.

[23] N. Hamdi and W. Chainbi, "A survey on energy aware vm consolidation strategies," *Sustainable Computing: Informatics and Systems*, vol. 23, pp. 80–87, 2019.

[24] Q. Zhou, M. Xu, S. S. Gill, C. Gao, W. Tian, C. Xu, and R. Buyya, "Energy efficient algorithms based on vm consolidation for cloud computing: Comparisons and evaluations," in *2020 20th IEEE/ACM International Symposium on Cluster, Cloud and Internet Computing (CCGRID)*, IEEE, 2020, pp. 489–498.

[25] X. Zhan and S. Reda, "Power budgeting techniques for data centers," *IEEE Transactions on Computers*, vol. 64, no. 8, pp. 2267–2278, 2015. DOI: 10.1109/TC.2014.2357810.

[26] D. G. Feitelson, *Workload modeling for computer systems performance evaluation*. Cambridge University Press, 2015.

[27] S. Khan, K. A. Shakil, and M. Alam, "Big data computing using cloud-based technologies, challenges and future perspectives," *arXiv preprint arXiv:1712.05233*, 2017.

[28] P. Carbone, A. Katsifodimos, S. Ewen, V. Markl, S. Haridi, and K. Tzoumas, "Apache flink: Stream and batch processing in a single engine," *Bulletin of the IEEE Computer Society Tech- nical Committee on Data Engineering*, vol. 36, no. 4, 2015.

[29] M. A. Netto, R. N. Calheiros, E. R. Rodrigues, R. L. Cunha, and R. Buyya, "Hpc cloud for scientific and business applications: Taxonomy, vision, and research challenges," *ACM Computing Surveys (CSUR)*, vol. 51, no. 1, pp. 1–29, 2018.

[30] I. Foster and C. Kesselman, *The Grid 2: Blueprint for a new computing infrastructure*. Elsevier, 2003.

[31] M. Armbrust, A. Fox, R. Griffith, A. D. Joseph, R. Katz, A. Konwinski, G. Lee, D. Patterson, A. Rabkin, I. Stoica, *et al.*, "A view of cloud computing," *Communications of the ACM*, vol. 53, no. 4, pp. 50–58, 2010.

[32] R. Buyya, C. Yeo, S. Venugopal, J. Broberg, and I. Brandic, "Cloud computing and emerging it platforms: Vision, hype, and reality for start [connect central controller using star topology network delivering computing as the 5th utility," *Future Generation Computer Systems*, vol. 25, pp. 599–616, 2009.

[33] E. K. Lee, H. Viswanathan, and D. Pompili, "Proactive thermal-aware resource management in virtualized hpc cloud datacenters," *IEEE Transactions on Cloud Computing*, vol. 5, no. 2, pp. 234–248, 2015.

[34] Q. Fang, J. Wang, Q. Gong, and M. Song, "Thermal-aware energy management of an hpc data center via two-time-scale control," *IEEE Transactions on Industrial Informatics*, vol. 13, no. 5, pp. 2260–2269, 2017.

[35] S. Malik and F. Huet, "Adaptive fault tolerance in real time cloud computing," in *2011 IEEE World Congress on services*, IEEE, 2011, pp. 280–287.

[36] J. A. Stankovic, "Misconceptions about real-time computing: A serious problem for next-generation systems," *Computer*, vol. 21, no. 10, pp. 10–19, 1988.

[37] K. H. Kim, A. Beloglazov, and R. Buyya, "Power-aware provisioning of cloud resources for real-time services," in *Proceedings of the 7th International Workshop on Middleware for Grids, Clouds and e-Science*, 2009, pp. 1–6.

[38] R. Sudeepa and H. Guruprasad, "Resource allocation in cloud computing," *International journal of modern communication technologies and research*, vol. 2, no. 4, p. 265 808, 2014.

[39] S. Esfandiarpoor, A. Pahlavan, and M. Goudarzi, "Structure-aware online virtual machine consolidation for datacenter energy improvement in cloud computing," *Computers & Electrical Engineering*, vol. 42, pp. 74–89, 2015.

[40] C. Pahl, "Containerization and the paas cloud," *IEEE Cloud Computing*, vol. 2, no. 3, pp. 24–31, 2015.

[41] P. Arroba, J. M. Moya, J. L. Ayala, and R. Buyya, "Dynamic voltage and frequency scaling-aware dynamic consolidation of virtual machines for energy efficient cloud data centers," *Concurrency and Computation: Practice and Experience*, vol. 29, no. 10, e4067, 2017.

[42] V. M. Raj and R. Shriram, "Power management in virtualized datacenter–a survey," *Journal of Network and Computer applications*, vol. 69, pp. 117–133, 2016.

[43] A. Al-Dulaimy, W. Itani, A. Zekri, and R. Zantout, "Power management in virtualized data centers: State of the art," *Journal of Cloud Computing*, vol. 5, no. 1, p. 6, 2016.

[44] A. Beloglazov, "Energy-efficient management of virtual machines in data centers for cloud computing," Ph.D. dissertation, 2013.

[45] H. David, C. Fallin, E. Gorbatov, U. R. Hanebutte, and O. Mutlu, "Memory power management via dynamic voltage/frequency scaling," in *Proceedings of the 8th ACM international conference on Autonomic computing*, 2011, pp. 31–40.

[46] Q. Deng, D. Meisner, A. Bhattacharjee, T. F. Wenisch, and R. Bianchini, "Coscale: Coordinating cpu and memory system dvfs in server systems," in *2012 45th annual IEEE/ACM international symposium on microarchitecture*, IEEE, 2012, pp. 143–154.

[47] A. Mahajan [1] and A. Ganpati, "A study of energy efficiency techniques in cloud computing," *International Journal of Computer Science and Mobile Computing*, vol. 3, no. 8, 2014.

[48] F. D. Rossi, M. G. Xavier, C. A. De Rose, R. N. Calheiros, and R. Buyya, "E-eco: Performance-aware energy-efficient cloud data center orchestration," *Journal of Network and Computer Applications*, vol. 78, pp. 83–96, 2017.

[49] M. G. Xavier, F. D. Rossi, C. A. De Rose, R. N. Calheiros, and D. G. Gomes, "Modeling and simulation of global and sleep states in acpi-compliant energy-efficient cloud environments," *Concurrency and Computation: Practice and Experience*, vol. 29, no. 4, e3839, 2017.

[50] C. Pahl, A. Brogi, J. Soldani, and P. Jamshidi, "Cloud container technologies: A state-of-the-art review," *IEEE Transactions on Cloud Computing*, 2017.

[51] I. Jimenez, C. Maltzahn, A. Moody, K. Mohror, J. Lofstead, R. Arpaci-Dusseau, and A. Arpaci-Dusseau, "The role of container technology in reproducible computer systems research," in *2015 IEEE International Conference on Cloud Engineering*, IEEE, 2015, pp. 379–385.

[52] R. Kumar and S. Charu, "An importance of using virtualization technology in cloud computing," *Global Journal of Computers & Technology*, vol. 1, no. 2, 2015.

[53] E. Arianyan, H. Taheri, and V. Khoshdel, "Novel fuzzy multi objective dvfs-aware consolidation heuristics for energy and sla efficient resource management in cloud data centers," *Journal of Network and Computer Applications*, vol. 78, pp. 43–61, 2017.

[54] T. Kaur and I. Chana, "Energy efficiency techniques in cloud computing: A survey and taxonomy," *ACM computing surveys (CSUR)*, vol. 48, no. 2, pp. 1–46, 2015.

[55] H. Yamada, "Survey on mechanisms for live virtual machine migration and its improvements," *Information and Media Technologies*, vol. 11, pp. 101–115, 2016.

[56] N. J. Kansal and I. Chana, "Energy-aware virtual machine migration for cloud computing-a firefly optimization approach," *Journal of Grid Computing*, vol. 14, no. 2, pp. 327–345, 2016.

[57] A.-Y. Son and E.-N. Huh, "Migration method for seamless service in cloud computing: Survey and research challenges," in *2016 30th International Conference on Advanced Information Networking and Applications Workshops (WAINA)*, IEEE, 2016, pp. 404–409.

[58] M. Liaqat, S. Ninoriya, J. Shuja, R. W. Ahmad, and A. Gani, "Virtual machine migration enabled cloud resource management: A challenging task," *arXiv preprint arXiv:1601.03854*, 2016.

[59] H. Liu, H. Jin, C.-Z. Xu, and X. Liao, "Performance and energy modeling for live migration of virtual machines," *Cluster computing*, vol. 16, no. 2, pp. 249–264, 2013.

[60] S. Akram, S. Ghaleb, S. Ba, and V. Siva, "Survey study of virtual machine migration techniques in cloud computing," *Int. J. Comput. Appl*, vol. 177, pp. 18–22, 2017.

[61] S. Sharma and M. Chawla, "A technical review for efficient virtual machine migration," in *2013 International Conference on Cloud & Ubiquitous Computing & Emerging Technologies*, IEEE, 2013, pp. 20–25.

[62] H. Liu, H. Jin, X. Liao, C. Yu, and C.-Z. Xu, "Live virtual machine migration via asynchronous replication and state synchronization," *IEEE Transactions on Parallel and Distributed Systems*, vol. 22, no. 12, pp. 1986–1999, 2011.

[63] D. Kapil, E. S. Pilli, and R. C. Joshi, "Live virtual machine migration techniques: Survey and research challenges," in *2013 3rd IEEE International Advance Computing Conference (IACC)*, IEEE, 2013, pp. 963–969.

[64] G. W. Dunlap, S. T. King, S. Cinar, M. A. Basrai, and P. M. Chen, "Revirt: Enabling intrusion analysis through virtual-machine logging and replay," *ACM SIGOPS Operating Systems Review*, vol. 36, no. SI, pp. 211–224, 2002.

[65] Ò. R. del Rıo, "Master of science in advanced mathematics and mathematical engineering,"

[66] T. Chaabouni and M. Khemakhem, "Energy management strategy in cloud computing: A perspective study," *The Journal of Supercomputing*, vol. 74, no. 12, pp. 6569–6597, 2018.

[67] M. C. Silva Filho, C. C. Monteiro, P. R. Inácio, and M. M. Freire, "Approaches for optimizing virtual machine placement and migration in cloud environments: A survey," *Journal of Parallel and Distributed Computing*, vol. 111, pp. 222–250, 2018.

[68] F. Farahnakian, T. Pahikkala, P. Liljeberg, J. Plosila, and H. Tenhunen, "Utilization prediction aware vm consolidation approach for green cloud computing," in *2015 IEEE 8th International Conference on Cloud Computing*, IEEE, 2015, pp. 381–388.

[69] M. Gamsız and A. H. Özer, "An energy-aware combinatorial virtual machine allocation and placement model for green cloud computing," *IEEE Access*, vol. 9, pp. 18 625–18 648, 2021.

[70] E. Arianyan, H. Taheri, and S. Sharifian, "Novel energy and sla efficient resource management heuristics for consolidation of virtual machines in cloud data centers," *Computers & Electrical Engineering*, vol. 47, pp. 222–240, 2015.

[71] ——, "Novel heuristics for consolidation of virtual machines in cloud data centers using multi-criteria resource management solutions," *The Journal of Supercomputing*, vol. 72, no. 2, pp. 688–717, 2016.

[72] ——, "Multi target dynamic vm consolidation in cloud data centers using genetic algorithm.," *Journal of Information Science & Engineering*, vol. 32, no. 6, 2016.

[73] E. Arianyan, H. Taheri, S. Sharifian, and M. Tarighi, "New six-phase on-line resource management process for energy and sla efficient consolidation in cloud data centers.," *Int. Arab J. Inf. Technol.*, vol. 15, no. 1, pp. 10–20, 2018.

[74] A. Aryania, H. S. Aghdasi, and L. M. Khanli, "Energy-aware virtual machine consolidation algorithm based on ant colony system," *Journal of Grid Computing*, vol. 16, no. 3, pp. 477–491, 2018.

[75] T. H. Duong-Ba, T. Nguyen, B. Bose, and T. T. Tran, "A dynamic virtual machine placement and migration scheme for data centers," *IEEE Transactions on Services Computing*, 2018.

[76] H. Li, W. Li, H. Wang, and J. Wang, "An optimization of virtual machine selection and placement by using memory content similarity for server consolidation in cloud," *Future Generation Computer Systems*, vol. 84, pp. 98–107, 2018.

[77] S. Mustafa, K. Bilal, S. U. R. Malik, and S. A. Madani, "Sla-aware energy efficient resource management for cloud environments," *IEEE Access*, vol. 6, pp. 15 004–15 020, 2018.

[78] F. Teng, L. Yu, T. Li, D. Deng, and F. Magoulès, "Energy efficiency of vm consolidation in iaas clouds," *The Journal of Supercomputing*, vol. 73, no. 2, pp. 782–809, 2017.

[79] F.-H. Tseng, X. Wang, L.-D. Chou, H.-C. Chao, and V. C. Leung, "Dynamic resource prediction and allocation for cloud data center using the multiobjective genetic algorithm," *IEEE Systems Journal*, vol. 12, no. 2, pp. 1688–1699, 2017.

[80] A. N. Toosi, C. Qu, M. D. de Assunção, and R. Buyya, "Renewable-aware geographical load balancing of web applications for sustainable data centers," *Journal of Network and Computer Applications*, vol. 83, pp. 155–168, 2017.

[81] H. Wang and H. Tianfield, "Energy-aware dynamic virtual machine consolidation for cloud datacenters," *IEEE Access*, vol. 6, pp. 15 259–15 273, 2018.

[82] M. Zakarya, "An extended energy-aware cost recovery approach for virtual machine migration," *IEEE Systems Journal*, vol. 13, no. 2, pp. 1466–1477, 2018.

[83] G. L. Stavrinides and H. D. Karatza, "Energy-aware scheduling of real-time workflow applications in clouds utilizing dvfs and approximate computations," in *2018 IEEE 6th International Conference on Future Internet of Things and Cloud (FiCloud)*, IEEE, 2018, pp. 33–40.

[84] A. Borghesi, A. Bartolini, M. Lombardi, M. Milano, and L. Benini, "Scheduling-based power capping in high performance computing systems," *Sustainable Computing: Informatics and Systems*, vol. 19, pp. 1–13, 2018.

[85] S. Mazumdar and M. Pranzo, "Power efficient server consolidation for cloud data center," *Future Generation Computer Systems*, vol. 70, pp. 4–16, 2017.

[86] A. Forestiero, C. Mastroianni, M. Meo, G. Papuzzo, and M. Sheikhalishahi, "Hierarchical approach for efficient workload management in geo-distributed data centers," *IEEE Transactions on Green Communications and Networking*, vol. 1, no. 1, pp. 97–111, 2016.

[87] A. Khosravi, L. L. Andrew, and R. Buyya, "Dynamic vm placement method for minimizing energy and carbon cost in geographically distributed cloud data centers," *IEEE Transactions on Sustainable Computing*, vol. 2, no. 2, pp. 183–196, 2017.

[88] N. Hogade, S. Pasricha, H. J. Siegel, A. A. Maciejewski, M. A. Oxley, and E. Jonardi, "Minimizing energy costs for geographically distributed heterogeneous data centers," *IEEE Transactions on Sustainable Computing*, vol. 3, no. 4, pp. 318–331, 2018.

[89] A. Khosravi, A. Nadjaran Toosi, and R. Buyya, "Online virtual machine migration for renewable energy usage maximization in geographically distributed cloud data centers," *Concurrency and Computation: Practice and Experience*, vol. 29, no. 18, e4125, 2017.

[90] H. Teyeb, N. B. Hadj-Alouane, S. Tata, and A. Balma, "Optimal dynamic placement of virtual machines in geographically distributed cloud data centers," *International Journal of Cooperative Information Systems*, vol. 26, no. 03, p. 1 750 001, 2017.

[91] N. Gholipour, E. Arianyan, and R. Buyya, "A novel energy-aware resource management technique using joint vm and container consolidation approach for green computing in cloud data centers," *Simulation Modelling Practice and Theory*, vol. 104, p. 102 127, 2020.

[92] M. Xu and R. Buyya, "Brownoutcon: A software system based on brownout and containers for energy-efficient cloud computing," *Journal of Systems and Software*, vol. 155, pp. 91–103, 2019.

[93] M. Rodriguez and R. Buyya, "Container orchestration with cost-efficient autoscaling in cloud computing environments," in *Handbook of research on multimedia cyber security*, IGI Global, 2020, pp. 190–213.

[94] N. Kumar, G. S. Aujla, S. Garg, K. Kaur, R. Ranjan, and S. K. Garg, "Renewable energy-based multi-indexed job classification and container management scheme for sustainability of cloud data centers," *IEEE Transactions on Industrial Informatics*, vol. 15, no. 5, pp. 2947–2957, 2018.

[95] H. Nie, P. Li, H. Xu, L. Dong, J. Song, and R. Wang, "Research on optimized pre-copy algorithm of live container migration in cloud environment," in *International Symposium on Parallel Architecture, Algorithm and Programming*, Springer, 2017, pp. 554–565.

[96] A. A. Khan, M. Zakarya, R. Buyya, R. Khan, M. Khan, and O. Rana, "An energy and performance aware consolidation technique for containerized datacenters," *IEEE Transactions on Cloud Computing*, 2019.

[97] A. Bhardwaj and C. R. Krishna, "A container-based technique to improve virtual machine migration in cloud computing," *IETE Journal of Research*, pp. 1–16, 2019.

[98] P. S. Junior, D. Miorandi, and G. Pierre, "Stateful container migration in geo-distributed environments," in *CloudCom 2020-12th IEEE International Conference on Cloud Computing Technology and Science*, 2020.

[99] M. A. Oxley, E. Jonardi, S. Pasricha, A. A. Maciejewski, H. J. Siegel, P. J. Burns, and G. A. Koenig, "Rate-based thermal, power, and co-location aware resource management for heterogeneous data centers," *Journal of Parallel and Distributed Computing*, vol. 112, pp. 126–139, 2018.